\documentclass[twoside]{article}
\usepackage{fleqn,espcrc2}

\newcommand{\be}{\begin{equation}}
\newcommand{\ee}{\end{equation}}
\newcommand{\beq}{\begin{eqnarray}}
\newcommand{\eeq}{\end{eqnarray}}
\newcommand{\nn}{\nonumber\\}
\newcommand{\non}{\nonumber}
\newcommand{\rpar}{\stackrel{\leftarrow}{\partial}}
\newcommand{\lpar}{\stackrel{\rightarrow}{\partial}}
\newcommand{\AmS}{{\protect\the\textfont2
  A\kern-.1667em\lower.5ex\hbox{M}\kern-.125emS}}

\title{Supersymmetric Model with Grassmann-odd Lagrangian}

\author{D.V. Soroka\address{Apt. 48, Val'ter Street 7, 61108 Kharkov,
        Ukraine},
	V.A. Soroka\address{Institute for Theoretical Physics, NSC
        "Kharkov Institute of Physics and Technology",\\
        61108 Kharkov, Ukraine}
	and
	J. Wess\address{Sektion Physik, Universit\"at M\"unchen,
        Theresienstr.\ 37, D-80333 M\"unchen}
	\address{Max-Planck-Institut f\"ur Physik,
        F\"ohringer Ring 6, D-80805 M\"unchen}}

\begin{document}

\begin{abstract}
A supersymmetric $D = 1, N =1$ model with a Grassmann-odd Lagrangian is
proposed which, in contrast to the model with an even Lagrangian, contains
not only a kinetic term but also an interaction term for the coordinates
entering into one real scalar Grassmann-even (bosonic) superfield.
\vspace{1pc}
\end{abstract}

% typeset front matter (including abstract)
\maketitle

\section{Introduction}

There are two ways for the formulation of Hamiltonian dynamics in
phase superspace: either by means of an even Poisson bracket with the help
of a Grassmann-even Hamiltonian or in an odd Poisson bracket \cite{l} with
a Grassmann-odd Hamiltonian\footnote{We shall only consider non-degenerate
brackets.}. The dynamical systems with equal numbers of pairs of even and
odd phase coordinates admit equivalent descriptions in the both brackets
at once \cite{vpst,s} with the help of the corresponding equivalent
Hamiltonians of opposite Grassmann parities. However, there are such
Hamiltonian systems which permit for their description only the bracket of
a definite parity. For example, the Hamilton systems with odd equal
numbers of the even and odd canonical variables can be described only by
means of the odd Poisson bracket (see, for example, \cite{vty}). On the
other hand, the systems having an even number of the even canonical
variables which differs from a number of the odd variables allow their
Hamiltonian description only in terms of the even Poisson bracket. In this
connection, by assuming that a Hamilton system has the corresponding to
it Lagrangian formulation and taking into account that in this case the
Grassmann parity of the Lagrangian coincides with the parities of the
bracket and Hamiltonian in terms of which this Hamiltonian system is
described, we see that there are also two ways for the formulation of
Lagrangian dynamics in configuration superspace:  either by means of
an even Lagrangian or in terms of an odd one \cite{s,f,s3}. As in the
case of the Hamiltonian dynamics, there are such systems which can be
described only with the help of the Grassmann-odd Lagrangian and have no
description in terms of the even one and vice versa.

In this report we just give such examples for $D = 1, N = 1$
supersymmetric models.  By using the formalism with the even Lagrangian,
we can construct from a real bosonic scalar superfield only kinetic
terms, but a term for interaction turns out to be a total time derivative.
While, using the formalism with the odd Lagrangian, we are able to
construct from this superfield a model which contains both the kinetic and
interaction terms.

The report is organized as follows. In Section 2 we show that in the case
$D =1, N = 1$ supersymmetry the even Lagrangian, which reduced from
Lagrangian density constructed from one real scalar bosonic superfield,
does not contain an interaction term. Then we propose a $D=1, N=1$
supersymmetric model having the odd Lagrangian, which also constructed
on the basis of this superfield but has an interaction term.
Consideration in this Section is carried out in the Lagrangian component
formulation. The Hamiltonian component formulation for this model is given
in Section 3. In Section 4 we present a superfield approach for the model
both in the Lagrangian and Hamiltonian formalisms. In Section 5 we show
once again that even in a particular case of the supersymmetric
one-dimensional oscillator the even Lagrangians, which are dynamically
equivalent to the model proposed, can not be reduced from any superfield
Lagrangian densities constructed from one real bosonic $D=1, N=1$
superfield.

\section{Lagrangian component formulation}

With respect to $D= 1, N = 1$ supersymmetry for the proper time $t$
and its real Grassmann superpartner $\eta$
\beq
t' = t + i \epsilon \eta\ ,\qquad \eta ' = \eta + \epsilon
\non\eeq
the components of the real Grassmann-even (bosonic)
($g(\Phi) = 0$)\footnote{$g(\Phi)$ is a Grassmann parity of the
quantity $\Phi$.} scalar superfield
\be\label{2.1}
\Phi(t, \eta) = q(t) + i \eta \theta(t) = {\Phi}'(t', \eta ')
\ee
are transformed as
\beq
\delta q(t) \equiv q'(t) - q(t) = i \theta \epsilon\ ,
\non\eeq
\be\label{2.2}
\delta \theta(t) \equiv \theta'(t) - \theta(t) = \dot q \epsilon\ ,
\ee
where a dot means a time derivative and $\delta$ is a variation of
the form.

With the help of a covariant derivative
\be\label{2.3}
D = \partial_\eta - i \eta \partial_t\ ,
\ee
where $\partial_x \equiv {\partial \over \partial x}$, we can construct
from $\Phi$ an invariant action $S_0$ with a Grassmann-even Lagrangian
${\stackrel{0}{l}}, (g({\stackrel{0}{l}}) = 0)$
\beq
S_0 = \int dt d\eta \left[{i \over 2} D \Phi \dot \Phi +
i D V(\Phi) \right] = \int dt {\stackrel{0}{l}}\ ,
\non\eeq
which contains only kinetic terms for the bosonic $q$ and fermionic
$\theta$ coordinates
\beq
{\stackrel{0}{l}} = {{\dot q}^2 \over 2} + {i \over 2}
\theta \dot \theta + {\dot V}(q)\ ,
\non\eeq
while their interaction term with an arbitrary real function
$V(\Phi)$ of $\Phi$ is a total time derivative.

By using a Grassmann-odd Lagrangian ${\stackrel{1}{L}}$
($g({\stackrel{1}{L}}) = 1$) which in the case of
$D = 1, N = 1$ supersymmetry corresponds to an even Lagrangian density
$\cal L$ ($g({\cal L}) = 0$), we are able to construct a supersymmetric
action
\beq\label{2.4}
S_1 = - i \int dt d\eta \left[ {{\dot \Phi}^2 \over 2} -
V(\Phi) \right] = - i \int dt d\eta {\cal L}\nn =
\int dt {\stackrel{1}{L}}\ ,
\eeq
including both a kinetic term for coordinates $q$, $\theta$ and a
term describing their interaction with the help of an arbitrary
real function $V(\Phi)$. The Lagrangian density superfield has the form
\be\label{2.5}
{\cal L} = {\stackrel{0}{L}} + i \eta{\stackrel{1}{L}} =
{{\dot q}^2 \over 2} - V(q) + i \eta [\dot q \dot \theta -
\theta V'(q) ]\ ,
\ee
where a prime denotes a derivative with respect to the coordinate
$q$. Note that the even and odd Lagrangians ${\stackrel{0}{L}}$ and
${\stackrel{1}{L}}$ are connected with each other with the aid of the
exterior differential\footnote{We adopt the Grassmann parity of the
exterior differential to be equal to unit $g(dx) = g(x) + 1$.} $d$ that
accompanied by the following map $\lambda : dq \rightarrow \theta$
(see, for example, \cite{s3})
\beq
\lambda : d{\stackrel{0}{L}} \rightarrow  {\stackrel{1}{L}}
\non\eeq
and are transformed under supersymmetry (\ref{2.2}) as
\beq
\delta {\stackrel{0}{L}} = i {\stackrel{1}{L}} \epsilon\ ,\qquad
\delta {\stackrel{1}{L}} = \dot {{\stackrel{0}{L}}} \epsilon\ .
\non\eeq
In accordance with the scheme given in \cite{s3}, the Euler-Lagrange
equation for the even Lagrangian ${\stackrel{0}{L}}$ coincides with that
one for the odd Lagrangian ${\stackrel{1}{L}}$, which corresponds to the
variable $\theta$,
\beq\label{2.6}
{d \over dt} \left(\partial_{\dot\theta}{\stackrel{1}{L}} \right) -
\partial_{\theta}{\stackrel{1}{L}} =
{d \over dt} \left(\partial_{\dot q}{\stackrel{0}{L}} \right) -
\partial_q{\stackrel{0}{L}}\nn = \ddot q + V'(q) = 0\ ,
\eeq
while the Euler-Lagrange equation for ${\stackrel{1}{L}}$, corresponding
to the variable $q$, can be obtained by taking the exterior differential
from two last equations in (\ref{2.6}) and performing the map $\lambda$
\beq\label{2.7}
{d \over dt} \left(\partial_{\dot q}d{\stackrel{0}{L}} \right) -
\partial_qd{\stackrel{0}{L}} =
d\ddot q + dqV'' = 0\nn \stackrel{\lambda} \rightarrow
{d \over dt} \left(\partial_{\dot q}{\stackrel{1}{L}} \right) -
\partial_q{\stackrel{1}{L}} = \ddot \theta + \theta V'' = 0\ .
\eeq

\section{Hamiltonian component formulation}

The Lagrangians ${\stackrel{0}{L}}$ and ${\stackrel{1}{L}}$ in (\ref{2.5})
have the same even momentum
\beq
p = \partial_{\dot q}{\stackrel{0}{L}} =
\partial_{\dot\theta}{\stackrel{1}{L}} = \dot q\ ,
\non\eeq
whereas an odd momentum $\pi$ for ${\stackrel{1}{L}}$ is connected with
it by means of the exterior differential $d$ and the map $\lambda$
\beq
\lambda : dp \rightarrow \pi =
\partial_{\dot q}{\stackrel{1}{L}} = \dot \theta\ .
\non\eeq
Corresponding to the even Lagrangian ${\stackrel{0}{L}}$ an even
Hamiltonian
\be\label{3.1}
{\stackrel{0}{H}} = \dot q p - {\stackrel{0}{L}} = {p^2 \over 2} + V(q)
\ee
is also related with an odd Hamiltonian ${\stackrel{1}{H}}$, responding
to the odd Lagrangian ${\stackrel{1}{L}}$,
\be\label{3.2}
{\stackrel{1}{H}} = \dot q \pi + \dot \theta p - {\stackrel{1}{L}} =
p \pi+ \theta V'(q)
\ee
in a similar way
\beq
\lambda : d{\stackrel{0}{H}} \rightarrow {\stackrel{1}{H}}\ .
\non\eeq

In the case of Hamiltonian formulation, the scheme given in
Refs. \cite{s3,n} prescribe that the Hamilton equations
\be\label{3.3}
\dot q = p\ ,\qquad \dot p = - V'(q)
\ee
formulated for the even phase coordinates $x^i = (q, p)$ with the help
of the even Hamiltonian (\ref{3.1}) in the even Poisson bracket
\be\label{3.4}
\dot x^i = ( x^i, {\stackrel{0}{H}} )_0 \equiv
x^i \left( \rpar_{q} \lpar_{p} -
\rpar_{p} \lpar_{q} \right) {\stackrel{0}{H}}
\ee
coincide with those ones obtained for $x^i$ by using the odd Hamiltonian
(\ref{3.2}) in the odd bracket
\beq\label{3.5}
\dot x^i = ( x^i, {\stackrel{0}{H}} )_0 =
\{ x^i, {\stackrel{1}{H}} \}_1 \equiv
x^i ( \rpar_{q} \lpar_{\pi}\nn - \rpar_{\pi} \lpar_{q} +
\rpar_{\theta} \lpar_{p} - \rpar_{p} \lpar_{\theta} )
{\stackrel{1}{H}}\ ,
\eeq
while the equations
\be\label{3.6}
\dot \theta = \pi\ ,\qquad \dot \pi = - \theta V''\ ,
\ee
which are Hamiltonian equations for the odd phase coordinates
$\theta^i = (\theta, \pi)$ described in the odd bracket (\ref{3.5})
\beq
\dot \theta^i = \{ \theta^i, {\stackrel{1}{H}} \}_1
\non\eeq
can be obtained from (\ref{3.3}) with the use of the exterior differential
and the map $\lambda$
\beq
dx^i \stackrel{\lambda} \rightarrow \theta^i\ .
\non\eeq
In relations (\ref{3.4}) and (\ref{3.5}) ${\rpar}$ and ${\lpar}$ are right
and left derivatives, respectively.

Note that the even bracket (\ref{3.4}) for arbitrary functions $A$, $B$ of
canonical variables rewritten in terms of the proper time
\beq
t[{\stackrel{0}{H}}(q,p), q] =
\int {dq' \over\sqrt{2[{\stackrel{0}{H}}(q,p) -
V(q')]}}\bigg\vert_{q' = q}\nn + t_0[{\stackrel{0}{H}}(q,p)]
\non\eeq
(an arbitrary function $t_0$ corresponds to the choice of the proper time
origin) and its canonically conjugate ${\stackrel{0}{H}}$ (\ref{3.1})
takes the form
\be\label{3.4'}
( A, B )_0 =
A \left( \rpar_{t} \lpar_{\stackrel{0}{H}} -
\rpar_{\stackrel{0}{H}} \lpar_{t} \right) B
\ee
and given in the canonical variables $t, {\stackrel{0}{H}}, {\stackrel{1}{H}}$
and $\Theta$, where
\be\label{3.7}
dt \stackrel{\lambda} \rightarrow \Theta = {\theta \over p} +
{\stackrel{1}{H}}{\partial_{\stackrel{0}{H}}t({\stackrel{0}{H}}, q)}\ ,
\ee
the odd bracket (\ref{3.5}) for any functions $A$, $B$ is
\beq\label{3.5'}
\{ A, B \}_1 = A ( \rpar_{t} \lpar_{\stackrel{1}{H}} -
\rpar_{\stackrel{1}{H}} \lpar_{t}  +
\rpar_{\Theta} \lpar_{\stackrel{0}{H}}\nn -
\rpar_{\stackrel{0}{H}} \lpar_{\Theta} ) B\ .
\eeq

By using the time derivative and Hamiltonian equations (\ref{3.3}),
(\ref{3.6}), we obtain from the transformations (\ref{2.2}) of the
configuration superspace variables $q$ and $\theta$ the supersymmetry
transformation rules for the phase superspace coordinates $z^M = (x^i,
\theta^i)$
\beq
\delta q = i \theta \epsilon\ ,\qquad \delta p = i \pi \epsilon\ ,
\non\eeq
\beq
\delta \theta = \epsilon p\ ,
\qquad \delta \pi = - \epsilon V'\ ,
\non\eeq
which can be uniformly represented by means of the odd bracket (\ref{3.5})
\beq
\delta z^M = \{ z^M, {\stackrel{1}{Q}} \}_1 \epsilon
\non\eeq
with the help of an even supercharge
\be\label{3.8}
{\stackrel{1}{Q}} = {\stackrel{0}{H}} + i \pi \theta =
{\stackrel{0}{H}} + i {\stackrel{1}{H}} \Theta\ ,
\ee
where the last expression follows from (\ref{3.2}) and (\ref{3.7}).

The even supercharge (\ref{3.8}) together with the odd Hamiltonian
(\ref{3.2}) satisfies in the odd bracket (\ref{3.5'}) $N = 1$ superalgebra
relations
\beq
\{ {\stackrel{1}{Q}}, {\stackrel{1}{Q}} \}_1 = 2 i {\stackrel{1}{H}}\ ,\qquad
\{ {\stackrel{1}{Q}}, {\stackrel{1}{H}} \}_1 = 0\ .
\non\eeq
Note that an even covariant derivative has the form
\be\label{3.9}
{\stackrel{1}{D}} = {\stackrel{0}{H}} - i \pi \theta =
{\stackrel{0}{H}} - i {\stackrel{1}{H}} \Theta
\ee
and obeys in the odd bracket (\ref{3.5'}) the following permutation
relations:
\beq
\{ {\stackrel{1}{D}}, {\stackrel{1}{D}} \}_1 = - 2i{\stackrel{1}{H}}\ ,
\qquad \{ {\stackrel{1}{D}}, {\stackrel{1}{H}} \}_1 = 0\ ,
\non\eeq
\beq
\{ {\stackrel{1}{D}}, {\stackrel{1}{Q}} \}_1 = 0\ .
\non\eeq

\section{Superfield approach}

The Euler-Lagrange equation
\beq
{d \over dt} \left( \partial_{\dot\Phi}{\cal L} \right) -
\partial_{\Phi}{\cal L} = 0
\non\eeq
for the even Lagrangian density superfield
\be\label{4.1}
{\cal L} = {{\dot \Phi}^2 \over 2} - V(\Phi)\ ,
\ee
which follows from (\ref{2.4}), takes the form
\be\label{4.2}
\ddot\Phi + V'(\Phi) = 0\ .
\ee
The components of (\ref{4.2}) coincide with equations (\ref{2.6}) and
(\ref{2.7}).  From (\ref{4.1}) we can define a superfield of the momentum
density
\beq
{\cal P} = p + i \eta \pi = \partial_{\dot\Phi}{\cal L} =
\dot\Phi = \dot q + i \eta \dot \theta
\non\eeq
and a superfield for the Hamiltonian density
\be\label{4.3}
{\cal H} = {\stackrel{0}{H}} + i \eta {\stackrel{1}{H}} =
\dot \Phi {\cal P} - {\cal L} = {{\cal P}^2 \over 2} + V(\Phi)\ .
\ee

It is remarkable enough that on functions $A$ and $B$, depending on the
real bosonic superfield $\Phi$ and the momentum density superfield
$\cal P$, we are able to introduce an {\it even} Poisson bracket of the
following form
\beq
\{ A, B \}_0 = A \left( \rpar_{\Phi} \lpar_{\cal P} -
\rpar_{\cal P} \lpar_{\Phi} \right) B\ .
\non\eeq
By means of this bracket with the help of the Hamiltonian density
(\ref{4.3}), we can determine the following superfield Hamiltonian
equations
\beq
\dot \Phi = \{ \Phi, {\cal H} \}_0 = {\cal P}\ ,
\non\eeq
\beq
\dot {\cal P} = \{ {\cal P}, {\cal H} \}_0 = - V'(\Phi)\ ,
\non\eeq
which components give Hamiltonian equations (\ref{3.3}), (\ref{3.6}) for
the phase variables $z^M = (q,p;\theta,\pi)$.

\section{Conclusion}

In the simplest case of the one-dimensional supersymmetric oscillator
\be\label{5.1}
{\stackrel{1}{H}}_{\rm osc} = p \pi + a^2\theta q\ ,\qquad
\left( V_{\rm osc} = {{a^2 q^2}\over2} \right)\ ,
\ee
where $a$ is a constant of the inverse length dimension, the Hamilton
equations for the canonical coordinates $z^M = (q,p;\theta,\pi)$
\beq
\dot q = p\ ,\qquad \dot p = - a^2 q\ ,\qquad \dot\theta = \pi\ ,\qquad
\dot\pi = - a^2 \theta
\non\eeq
can be reproduced both with the odd Hamiltonian (\ref{5.1}) in the
canonical odd bracket (\ref{3.5}) and in the following even Dirac brackets
${\{...,...\}}^*_{0k}$, $(k = \pm1)$
\beq\label{5.2}
\dot z^M = \{z^M, {\stackrel{1}{H}}_{\rm osc} \}_1 =
{\{z^M, {\stackrel{0}{H}}_k \}}^*_{0k}\nn \equiv
z^M [ \rpar_{q} \lpar_{p} - \rpar_{p} \lpar_{q} - ik
( \rpar_{\theta} \lpar_{\theta}\nn +
\rpar_{\pi}a^2\lpar_{\pi} ) ] {\stackrel{0}{H}}_k
\eeq
with the aid of the corresponding even Hamiltonians
\be\label{5.3}
{\stackrel{0}{H}}_k = {{p^2 + a^2 q^2}\over2} - ik\pi\theta\ ,
\ee
which coincide in the case with the expressions for the even supercharge
${\stackrel{1}{Q}}_{\rm osc}$  (\ref{3.8}) and even covariant derivative
${\stackrel{1}{D}}_{\rm osc}$ (\ref{3.9})
\beq
{\stackrel{0}{H}}_{-1} = {\stackrel{1}{Q}}_{\rm osc}\ ,\qquad
{\stackrel{0}{H}}_1 = {\stackrel{1}{D}}_{\rm osc}\ .
\non\eeq

The even Dirac brackets (\ref{5.2}) and the even Hamiltonians (\ref{5.3})
follow from the even Lagrangians
\be\label{5.4}
{\stackrel{0}{L}}_k = {1\over 2}\ [ {\dot q}^2 - a^2 q^2 +
ik ( \psi^\alpha {\dot\psi}^\alpha + 2 a \psi^1 \psi^2 )]\ ,
\ee
where $\psi^\alpha(t)$ $(\alpha = 1,2)$ are two real fermionic
coordinates. Indeed, the even Lagrangians (\ref{5.4}) lead to the momenta
\be\label{5.5}
p = \partial_{\dot q}{\stackrel{0}{L}}_k = \dot q\ , \qquad
\pi_k^\alpha = \partial_{{\dot\psi}^\alpha}{\stackrel{0}{L}}_k =
- {ik \over 2} \psi^\alpha\ ,
\ee
canonically conjugate to the coordinates $q$ and $\psi^\alpha$ in the even
brackets corresponding to ${\stackrel{0}{L}}_k$
\be\label{5.6}
\{q, p\}_{0k} = 1, \qquad \{\psi^\alpha, \pi_k^\beta \}_{0k} =
- \delta^{\alpha \beta}\ .
\ee
The remaining even-bracket relations between the canonical variables have
zero right-hand sides. The last relations in (\ref{5.5}) define the
second-class constraints
\be\label{5.7}
\varphi_k^\alpha = \pi_k^\alpha + {ik \over 2} \psi^\alpha,\qquad
\{ \varphi_k^\alpha, \varphi_k^\beta \}_0 = - ik \delta^{\alpha\beta},
\ee
which commute in the even brackets (\ref{5.6}) with the variables
\beq
\chi_k^\alpha = \pi_k^\alpha - {ik \over 2} \psi^\alpha\ ,\qquad
\{ \chi_k^\alpha, \chi_k^\beta \}_0 = ik \delta^{\alpha\beta}\ ,
\non\eeq
\beq
\{ \varphi_k^\alpha, \chi_k^\beta \}_0 = 0\ ,
\non\eeq
entering into the definitions for the even Dirac brackets (\ref{5.2})
\beq\label{5.2'}
{\{z^M, {\stackrel{0}{H}}_k \}}^*_{0k} =
{\{z^M, {\stackrel{0}{H}}_k \}}_{0k}\nn -
ik \{z^M, \varphi_k^\alpha\}_{0k}
\{\varphi_k^\alpha, {\stackrel{0}{H}}_k \}_{0k}\nn =
z^M ( \rpar_{q} \lpar_{p} - \rpar_{p} \lpar_{q}\nn +
ik \rpar_{\chi_k^\alpha} \lpar_{\chi_k^\alpha} ) {\stackrel{0}{H}}_k
\eeq
and the even Hamiltonians (\ref{5.3})
\be\label{5.3'}
{\stackrel{0}{H}}_k = {{p^2 + a^2 q^2}\over2} + ika \chi_k^1\chi_k^2\ .
\ee
There is no summation over $k$ in the foregoing formulae.
The even Hamiltonians (\ref{5.3'}) follow from the total even Hamiltonians
corresponding to the Lagrangians (\ref{5.4}) with the use of the
second-class constraints $\varphi_k^\alpha = 0$. A correspondence between
different expressions (\ref{5.2}), (\ref{5.2'}) for the even Dirac
brackets and (\ref{5.3}), (\ref{5.3'}) for the even Hamiltonians
${\stackrel{0}{H}}_k$ can be established by putting either
\beq
a\chi_k^1 = \pm i\pi\ ,\qquad \chi_k^2 = \pm i\theta
\non\eeq
or
\beq
a\chi_k^2 = \pm i\pi\ ,\qquad \chi_k^1 = \mp i\theta\ .
\non\eeq

Note that even in the simplest case of the one-dimensional supersymmetric
oscillator the even Lagrangians (\ref{5.4}) do not follow from any
Lagrangian densities, given in terms of one real scalar bosonic
$D = 1, N = 1$ superfield $\Phi$ (\ref{2.1})\footnote{In the case $k = 1$
the even Lagrangian ${\stackrel{0}{L}}_1$ (\ref{5.4}) is reduced from  the
even Lagrangian density expressed in terms of one scalar bosonic $D = 1,~N
= 2$ superfield (see, for example, \cite{s3}).}, as it takes place for the
odd Lagrangian
\beq
{\stackrel{1}{L}}_{\rm osc} = \dot q\dot\theta - a^2 q\theta\ ,
\non\eeq
that leads to the odd Hamiltonian (\ref{5.1}), which is dynamically
equivalent (see equation (\ref{5.2}))  to the even one (\ref{5.3})
followed from the even Lagrangian (\ref{5.4}).

Thus, we see that in the case of $D = 1,~N = 1$ supersymmetry the
description with the odd Lagrangian is {\it more economical}.

\section{Acknowledgements}

One of the authors (V.A.S.) would like to thank the administration of the
Ludwig Maximilian University of M\"unchen for hospitality during the
period when the main part of this work has been done. V.A.S. is
also sincerely grateful to J. Lukierski for kind hospitality at the
University of Wroclaw where the part of the work has been performed and
together with A. Frydryszak for useful discussions.

\medskip
The research of V.A.S. was supported in part by the Ukrainian State
Foundation of Fundamental Researches, Grant No 2.5.1/54 and by Grant INTAS
No 93-127 (Extension).

\end{document}